\documentclass[fleqn,usenatbib]{mnras}

\usepackage[multidot]{grffile}

\usepackage[utf8]{inputenc}
\usepackage[T1]{fontenc}
\usepackage{ae,aecompl}

\DeclareRobustCommand{\VAN}[3]{#2}
\let\VANthebibliography\thebibliography
\def\thebibliography{\DeclareRobustCommand{\VAN}[3]{##3}\VANthebibliography}

\usepackage{graphicx}
\usepackage{amsmath}
\usepackage{amssymb}

\usepackage{comment}
\usepackage{color}
\usepackage{booktabs}
\usepackage{tabularx}
\usepackage{threeparttable}

\usepackage{academicons}
\usepackage{svg}
\newcommand{\orcid}[1]{\href{https://orcid.org/#1}{\textsuperscript{\includegraphics[width=10pt]{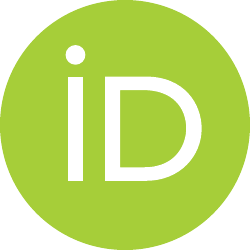}}}}

\newcommand{\mmode}[1]{\ifmmode{#1}\else{$#1$}\fi}

\newcommand{\kepler}[0]{\emph{Kepler}}

\newcommand{\Gaia}[0]{\emph{Gaia}}
\newcommand{\Teff}[0]{\mmode{T_\text{eff}}}
\newcommand{\Rsolar}[0]{\mmode{\text{R}_{\odot}}}
\newcommand{\Msolar}[0]{\mmode{\text{M}_{\odot}}}

\newcommand{\Dnu}[0]{\mmode{\Delta\nu}}
\newcommand{\dnu}[1]{\mmode{\delta\nu_{#1}}}
\newcommand{\numax}[0]{\mmode{\nu_\text{max}}}

\newcommand{\muHz}[0]{\mmode{\mu{\rm Hz}}}

\newcommand{\fDnu}[0]{\mmode{f_{\Delta\nu}}}

\newcommand{\dnum}[0]{\mmode{\delta\nu_{\rm m}}}

\newcommand{\Yinit}[0]{\mmode{Y_{\rm init}}}

\newcommand{\amlt}[0]{\mmode{\alpha_{\rm MLT}}}

\newcommand{\cyan}[1]{\textcolor{cyan} }

\usepackage{CJKutf8}
\newcommand{\CNnames}[1]{{\begin{CJK}{UTF8}{gbsn}~(#1)~\end{CJK}}}

\makeatletter
\newcommand\thefontsize[1]{{#1 The current font size is: \f@size pt\par}}
\makeatother

\defcitealias{sharma++2016-population-rg-kepler}{S16}

\title[Prescribing the surface correction]{A prescription for the asteroseismic surface correction}

\author[Y. Li et al.]{%
Yaguang Li\CNnames{李亚光}\orcid{0000-0003-3020-4437},$^{1}$ \thanks{yaguang.li@sydney.edu.au}
Timothy R. Bedding\orcid{0000-0001-5222-4661},$^{1}$ \thanks{tim.bedding@sydney.edu.au}
Dennis Stello\orcid{0000-0002-4879-3519},$^{2}$
Daniel Huber\orcid{0000-0001-8832-4488}$^{3}$ 
\newauthor
Marc Hon\orcid{0000-0003-2400-6960}$^{3}$,
Meridith Joyce\orcid{0000-0002-8717-127X},$^{4}$
Tanda Li\CNnames{李坦达}\orcid{0000-0001-6396-2563},$^{5}$
Jean Perkins\orcid{0000-0002-6703-5406},$^{6}$
Timothy R. White\orcid{0000-0002-6980-3392}$^{1}$,
\newauthor
Joel C. Zinn\orcid{0000-0002-7550-7151},$^{7}$
Andrew W. Howard\orcid{0000-0001-8638-0320},$^{8}$
Howard Isaacson\orcid{0000-0002-0531-1073},$^{9}$
Daniel R. Hey\orcid{0000-0003-3244-5357}$^{3}$ and
Hans Kjeldsen\orcid{0000-0002-9037-0018}$^{10}$
\\
$^{1}$Sydney Institute for Astronomy (SIfA), School of Physics, University of Sydney, NSW 2006, Australia\\
$^{2}$School of Physics, University of New South Wales, 2052, Australia\\
$^{3}$Institute for Astronomy, University of Hawai`i, 2680 Woodlawn Drive, Honolulu, HI 96822, USA\\
$^{4}$Space Telescope Science Institute, 3700 San Martin Dr, Baltimore, MD 21218, USA\\
$^{5}$Department of Astronomy, Beijing Normal University, Haidian District, Beijng 100875, China\\
$^{6}$Monterey Institute for Research in Astronomy, 200 8th St, Marina, CA 93933\\
$^{7}$Department of Astrophysics, American Museum of Natural History, Central Park West at 79th Street, New York, NY 10024, USA\\
$^{8}$Division of Physics, Mathematics and Astronomy, Caltech, 1200 E California Blvd, Pasadena CA 91125, USA\\
$^{9}$Department of Astronomy, University of California at Berkeley, 501 Campbell Hall, Berkeley, CA 94720-3411, USA\\
$^{10}$Stellar Astrophysics Centre, Department of Physics and Astronomy, Aarhus University, 
8000 Aarhus C, Denmark\\
}

\date{Accepted XXX. Received YYY; in original form ZZZ}

\pubyear{2021}

\begin{document}
\label{firstpage}
\pagerange{\pageref{firstpage}--\pageref{lastpage}}
\maketitle

\begin{abstract}
In asteroseismology, the surface effect refers to a disparity between the observed and the modelled frequencies in stars with solar-like oscillations. It originates from improper modelling of the surface layers.
Correcting the surface effect usually requires using functions with free parameters, which are conventionally fitted to the observed frequencies.
On the basis that the correction should vary smoothly across the H--R diagram, we parameterize it as a simple function of surface gravity, effective temperature, and metallicity.
We determine this function by fitting a wide range of stars.
The absolute amount of the surface correction decreases with luminosity, but the ratio between it and \numax{} increases, suggesting the surface effect is more important for red giants than dwarfs.
Applying the prescription can eliminate unrealistic surface correction, which improves parameter estimations with stellar modelling.
Using two open clusters, we found a reduction of scatter in the model-derived ages for each star in the same cluster.
As an important application, we provide a new revision for the \Dnu{} scaling relation that, for the first time, accounts for the surface correction.
The values of the correction factor, \fDnu{}, are up to 2\% smaller than those determined without the surface effect considered, suggesting decreases of up to 4\% in radii and up to 8\% in masses when using the asteroseismic scaling relations. 
This revision brings the asteroseismic properties into an agreement with those determined from eclipsing binaries.
The new correction factor and the stellar models with the corrected frequencies are available at {{https://www.github.com/parallelpro/surface}}.
\end{abstract}





\section{Introduction}
\label{sec:intro}

\begin{figure}
\includegraphics[width=\columnwidth]{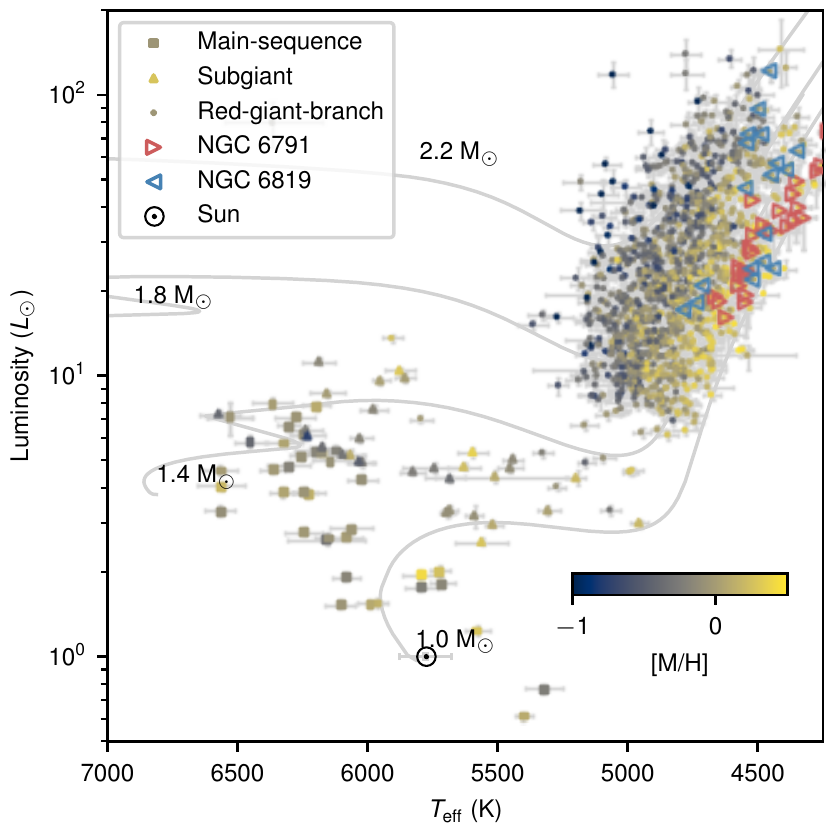}
\caption{H--R diagram showing the studied sample in this work. The evolutionary tracks for four masses with $Y_{\rm init}=0.29$, ${\rm [M/H]}=0.0$, $\alpha_{\rm MLT}=1.7$ are shown in grey lines. Note these model parameters are approximate, not exact, since the tracks were generated pseudo-randomly (see Sec.~\ref{subsec:model}). }
\label{fig:hrd}
\end{figure}

Correcting the asteroseismic surface effect has so far been a troublesome procedure. 
Convection affects pulsation properties through turbulent pressure, opacity variations, and convective energy flux \citep{houdek++2017-surface-physics-solar-frequency}. 
Small-scale magnetic fields can form layers that affect the propagation of pulsations \citep{liy+2021-surface-magnetic}.
All these processes are poorly modelled in the near-surface convective atmosphere in most 1D stellar models \citep{jcd++1988-eos,dziembowski++1988-surface}. 
Improvements have been seen with the surface layers replaced by 3D averaged atmospheric models, producing more realistic equilibrium structures
\citep{rosenthal++1999-solar-convection,magic+2016-solar-surf-corr,jorgensen++2017-frequency-solartype-stellar-model-3d-atmos,trampedach++2017-surface-effect-3d-simulation-1-convective-expansion-atmos,jorgensen++2018-3d-on-the-fly-1,jorgensen++2019-surf-corr-3d,mosumgaard++2020-3d-on-the-fly-2}, or with time-dependent 1D convection models, accounting for the coupling between oscillation and convection \citep{balmforth-1992-pulsational-stability-2-frequency,grigahcene++2012-surface,jcd-2012-stellar-models,houdek++2017-surface-physics-solar-frequency,houdek++2019-damping-rate-lagacy-kepler,belkacem++2021-surface-turbulence,philidet++2021-surface-1}.

In practice, the surface effect is usually corrected empirically with simple functions of frequency.
\citet{jcd++1989-differential-inversion} provided a justification, based on a perturbation to an asymptotic formalism of acoustic modes. 
By rescaling the frequency correction obtained from the solar standard model, \citet{silvaaguirre++2015-33-kepler-exoplanet-host} and \citet{houdek++2019-damping-rate-lagacy-kepler} applied it to other main-sequence stars.
Several other correction formula were also put forward \citep[e.g.][]{kjeldsen++2008-surface-correction,sonoi++2015-surf-corr-3d}.
In particular, \citet{gough-1990-seis-progress} suggested that the corrections are proportional to the cubic and the inverse of frequencies scaled by mode inertia:
\begin{equation}
\label{eq:inv-cubic}
\dnu{} = \left[a_{3} (\nu/\numax{})^3 + a_{-1} (\nu/\numax{})^{-1}\right]/I,
\end{equation} 
where $a_{3}$ and $a_{-1}$ are the free parameters to be determined. The frequency of maximum power, \numax{}, is evaluated via the scaling relation \citep{brown++1991-dection-procyon-scaling-relation,kjeldsen+1995-scaling-relations}:
\begin{equation}\label{eq:numax}
    \frac{\numax}{\nu_{{\rm max,\odot}}} \approx \frac{g}{g_\odot} \left(\frac{\Teff}{T_{{\rm eff},\odot}}\right)^{-1/2},
\end{equation}
where we adopt $g_\odot=274$ ${\rm m/s^2}$, $T_{\rm eff,\odot}=5777$ K, and $\nu_{\rm max,\odot}=3090$ \muHz{} throughout this work.
Since the cubic term usually dominates the frequency correction, another correction form is written as
\begin{equation}
\label{eq:cubic}
\dnu{} = a_{3} (\nu/\numax{})^3 /I,
\end{equation} 
where $a_{3}$ is the free parameter.

These two functional forms have shown to match observations quite well.
\citet{ball+2014-surface-correction-inertia-weighted,ball+2017-six-rg-surface-correction} showed that they work well for radial modes on the Sun and red-giant-branch stars, albeit with some caveats for mixed modes \citep{ong++2021-surf-corr-theory,ong++2021-surf-corr-sg}.
Many works concluded the inverse-cubic form could obtain an overall good fit  \citep{schimitt+2015-surf-corr-HR,compton++2018-surf-corr-legacy,nsamba++2018-input-physics,jorgensen++2020-surf-corr-method-comparison} and correctly recover the dynamical stellar properties of binary systems \citep{jorgensen++2020-surf-corr-method-comparison}.

The correction usually works as follows. 
Given a star with a set of observational frequencies and a stellar model with a set of theoretical frequencies, one can calculate the difference between the two frequency sets. 
This difference is then fitted to the right-hand-side of the frequency correction function (Eq.~\ref{eq:inv-cubic} or~\ref{eq:cubic}) to determine the free parameters. 
The amount of frequency correction is then calculated with the best-fitting values and added to the theoretical frequencies. 

One problem with this method is that the surface correction can only be determined with a fit to observed frequencies. 
It does not allow us to estimate the surface terms for any theoretical model without being close to the observed star. 
More seriously, it can lead to a model with an unphysically large (or small) surface correction that fits the data well but is a poor representation of the star. 
In this paper, we tackle these problems through a simple prescription for the surface effect, assuming that it varies smoothly with stellar parameters (Sec.~\ref{sec:prescription}).
This variation is then constrained by an ensemble fit to a wide range of stars (Sec.~\ref{sec:sample}). 
Adopting this prescription improves parameter estimations with stellar modelling (Sec.~\ref{sec:params}). 
It further enables an improved correction to the commonly-used \Dnu{} scaling relation (Sec.~\ref{sec:dnu}). 

\section{Prescribing the surface correction}
\label{sec:prescription}




Since the surface effect originates from the model atmosphere, it is reasonable to assume it is a smooth function of surface parameters, i.e. surface gravity $g$, effective temperature \Teff{}, and metallicity [M/H]. 
This assumption is supported by 3D atmospheric simulations \citep{sonoi++2015-surf-corr-3d,manchon++2018-surface-feh} and 1D non-adiabatic convection models \citep{houdek++2019-damping-rate-lagacy-kepler}. 
These works suggested that the surface correction at \numax{}, denoted by \dnum{}, varies from star to star as a function of \Teff{} and $g$.
Hence, we propose a prescription for \dnum{} as follows:
\begin{equation} 
\label{eq:dnum}
\dnum = a \cdot (g/{g_\odot})^{b} \cdot (T_{\rm eff}/T_{\rm eff,\odot})^{c} \cdot (d \cdot {\rm [M/H]} + 1),
\end{equation}
where the free parameters to be determined are $\theta_s=\{a,b,c,d\}$. 
By construction, the parameter $a$ is the amount of surface correction at \numax{} for a solar model.
If we adopt the cubic formula, for each star we can directly use Eq.~\ref{eq:dnum} to solve the surface term $a_{3}$ in Eq.~\ref{eq:cubic} with $\nu$ equal to \numax{}. 
To obtain the mode inertia $I$ on the RHS of Eq.~\ref{eq:cubic}, we interpolated $\nu^3/I$ to the frequency \numax{}.

If we adopt the inverse-cubic formula to correct model frequencies, another equation is needed since there are two surface terms, $a_{-1}$ and $a_{3}$.
We propose that the surface correction at $s$ times of \numax{}, denoted by $\delta\nu_{\rm m}'$, also varies with the surface parameters:
\begin{equation} 
\label{eq:dnumf}
\delta\nu_{m}' = a' \cdot (g/{g_\odot})^{b'} \cdot (T_{\rm eff}/T_{\rm eff,\odot})^{c'} \cdot (d' \cdot {\rm [M/H]} + 1).
\end{equation}
Together with Eq.~\ref{eq:dnum}, the free parameters in this prescription are $\theta_s=\{a,b,c,d,a',b',c',d'\}$.
By varying the value of $s$, we found no obvious changes to the solutions of those free parameters. 
Hence, we fixed $s$ at 1.1, so that $\delta\nu_{\rm m}'$ represents the amount of surface correction at $1.1\numax{}$. 
For each star, we then used Eq.~\ref{eq:dnum} and Eq.~\ref{eq:dnumf} to solve $a_{-1}$ and $a_{3}$ in Eq.~\ref{eq:inv-cubic} with $\nu=\numax$ and $\nu=1.1\numax$, respectively.
To calculate the RHS of Eq.~\ref{eq:inv-cubic}, we interpolated $\nu^3/I$ and $\nu^{-1}/I$ to the frequency \numax{}.




\begin{table*} 
\caption{Stellar parameters of the studied sample. 
\label{tab:stellar-params} \\ 
} 
\begin{tabular*}{1.01\textwidth}{@{\extracolsep{\fill}}lrrrrrrrrrrrr}
\toprule
       Star &   $L$ & $\sigma_L$ & Ref($L$) & $T_{\rm eff}$ & $\sigma_{T_{\rm eff}}$ & Ref($T_{\rm eff}$) & ${\rm [M/H]}$ & $\sigma_{\rm [M/H]}$ & Ref(${\rm [M/H]}$) \\
\midrule
  $\rm Sun$ &   1.0 &       0.02 &      --- &          5777 &                    100 &                --- &           0.0 &                 0.05 &                --- \\
  $\mu$ Her &  2.54 &       0.08 &        2 &          5560 &                    100 &                  3 &          0.28 &                 0.05 &                  3 \\
KIC10000547 & 12.68 &       0.51 &        0 &          4969 &                     36 &                  1 &         -0.26 &                 0.05 &                  6 \\
KIC10001440 & 39.77 &        3.8 &        0 &          4773 &                     42 &                  1 &         -0.65 &                 0.05 &                  6 \\
KIC10004825 & 42.35 &       3.34 &        0 &          4611 &                     55 &                  1 &          0.21 &                 0.05 &                  6 \\
KIC10014893 & 68.69 &       4.67 &        0 &          4579 &                     26 &                  1 &         -0.13 &                 0.05 &                  6 \\
KIC10014959 &  9.29 &       0.33 &        0 &          4813 &                     29 &                  1 &          0.11 &                 0.05 &                  6 \\
KIC10018442 &  37.7 &       1.89 &        0 &          4781 &                     41 &                  1 &         -0.03 &                 0.05 &                  6 \\
KIC10018811 & 27.11 &       1.41 &        0 &          4918 &                     28 &                  1 &         -0.29 &                 0.05 &                  6 \\
\bottomrule
\end{tabular*}
 \begin{tablenotes}  
 \item \emph{Note}: References for the stellar parameters: 0 (This work); 1 \citep{casagrande++2021-irfm-gaia}; 2 \citep{grundahl++2017-mu-her}; 3 \citep{jofre++2015-star-params}; 4 \citep{furlan++2018-kepler}; 5 \citep{lund++2017-legacy-kepler-1}; 6 \citep{ahunmada++2020-apogee-dr16}; 7 \citep{buchhave++2015-feh}. Only the first 10 lines are shown. The full table can be accessed online.
 \end{tablenotes}
\end{table*}

\section{Data analysis}
\label{sec:sample}

\subsection{Observational sample}
\label{subsec:obs} 

In order to constrain Eqs.~\ref{eq:dnum} and~\ref{eq:dnumf}, we need a sample of stars spanning a sufficiently large parameter space. 
Our sample (see Fig.~\ref{fig:hrd}) consists of stars with measured individual frequencies: the Sun \citep{broomhall++2009-bison-freqs}, the SONG subgiant $\mu$ Herculis \citep{grundahl++2017-mu-her}, \kepler{} main-sequence dwarfs \citep{lund++2017-legacy-kepler-1}, \kepler{} subgiants \citep{liyg++2020-kepler-36-subgiants} and \kepler{} red-giant-branch (RGB) stars with $\Dnu>~2$~$\mu$Hz \citep{litd++2022-kepler-rgb}. 
The RGB stars were distinguished from helium-burning stars by \citet{bedding++2011-distinguish-rc-rgb}, \citet{stello++2013-13000-rg-kepler}, \citet{mosser++2014-mixed-mode-rg-window}, \citet{vrard++2016-period-spacings}, \citet{elsworth-2017-asteroseismic-evolution-state}, and \citet{hon++2017-deep-learning-rc-rgb}. 

We compiled metallicities [M/H] from various sources, including HIRES spectra (see below), APOGEE DR16 \citep{ahunmada++2020-apogee-dr16}, \citet{lund++2017-legacy-kepler-1}, and \citet{buchhave++2015-feh} (listed in the order of priority) wherever possible. 
We collected metallicities for 36 stars measured with HIRES spectrograph \citep{vogt94} at the Keck-I 10-m telescope on Maunakea observatory, Hawai`i by \citet{furlan++2018-kepler}. We also obtained new HIRES spectra for 21 stars in this work. The spectra were obtained and reduced as part of the California Planet Search queue \citep[CPS,][]{howard10}. We used the C5 decker and obtained spectra with a S/N per pixel of 80 at $\sim 600$\,nm with a spectral resolving power of $R \sim 60 000$. To measure the metallicities, we applied Specmatch-synth \citep{petigura15}, which fits a synthetic grid of model atmospheres and has been extensively validated through the California Kepler Survey \citep{petigura17,johnson17}. 
All metallicity measurements were brought to the APOGEE abundance scale by adding constant offsets, determined with the [M/H] measurements of same stars. Because of the limited number of metal-poor stars, we restricted our sample to have $\text{[M/H]}>-0.8$ dex.

We determined the effective temperatures, \Teff{}, with \Gaia{} and 2MASS photometry, using the infrared flux method (IRFM) calibrated by \citet{casagrande++2021-irfm-gaia}. This \Teff{} scale was benchmarked against solar twins, Gaia benchmark stars, and interferometry.

We determined luminosities, $L$, using \Gaia{} DR3 \citep{gaia-2016-mission,gaia-2020-edr3}. \Gaia{} parallaxes are known to have zero-point offsets, which we corrected using a model from \citet{lindegren++2020-gaia-edr3-parallax-zpt}. The reported parallaxes also have underestimated uncertainties. Therefore we inflated them by a factor of 1.3, according to external calibrations \citep{elbadry++2021-binary-gaia-dr3,zinn-2021-gaia-dr3-plx-seismology,maz-apellaniz++2021-gaia-dr3-plx-cluster}. 
We then calculated the luminosities by combining the parallaxes with the 2MASS $K$-band magnitudes and using the ``direct'' method in the software {\small ISOCLASSIFY} \citep{huber++2017-seismic-radii-gaia,berger++2020-gaia-kepler-1-stars}, which implements the \cite{green++2019-dustmap} dust map and the bolometric corrections from MIST models \citep{Choi++2016-mist-1-solar-scaled-models}.

Additionally, we used RGB stars from two \kepler{} clusters as a test sample: NGC 6791 \citep{basu++2011-model-ngc6791-ngc6819,mckkever++2019-ngc6791,brogaard++2021-ngc6791} and NGC 6819 \citep{stello++2010-ngc6819,corsaro++2012-kepler-clusters,handberg++2017-ngc6819}. These cluster stars were not used for fitting the prescription, but for validating the result (Sec.~\ref{sec:params}). 
We estimated their stellar parameters following the same procedure illustrated above. 
Table~\ref{tab:stellar-params} lists the stellar parameters used in modelling.
Fig.~\ref{fig:hrd} shows an overview of our sample on the H--R diagram.


\subsection{Stellar models}
\label{subsec:model} 
We calculated a grid of stellar models using Modules for Experiments in Stellar Astrophysics \citep[{\small MESA}, version r15140;][]{paxton++2011-mesa,paxton++2013-mesa,paxton++2015-mesa,paxton++2018-mesa,paxton++2019-mesa} to model stellar evolution and structure, and {\small GYRE} \citep[version 6.0.1;][]{townsend+2013-gyre} to calculate adiabatic frequencies from the structure profiles computed from {\small MESA}.

Here, we summarise the input physics for the constructed models. 
We used the \citet{henyey++1965-mlt} description of the mixing length theory to formulate convection, with the mixing length being one of the free parameters, since a solar-calibrated mixing length can not fit stars with various stellar properties \citep{tayar-2017-mixing-length-metallicity-rg-age-gaia,joyce+2018-amlt-metal-poor}. 
We set the convective overshoot with an exponential scheme discussed by \citet{herwig-2000-overshoot}.
For core overshoot, we set the efficiency parameter $f_{\rm ov, core}$ as a function of mass, according to the calibration from eclipsing binaries \citep[equation 2 of ][]{claret++2018-eb-core-fov}.
For envelope overshoot, we set $f_{\rm ov, env}$ as 0.006, according to a solar calibration with our adopted input physics.

We chose the current solar photospheric abundance as the reference scale for metallicity: $X_{\odot}=0.7381$, $Y_{\odot}=0.2485$, $Z_{\odot}=0.0134$ \citep[][the AGSS09 scale]{asplund++2009-solar-composition-review}. 
Hence the metallicity is
\begin{equation} 
{\rm [M/H]} = \log_{10}(Z/X) - \log_{10}(Z_{\odot}/X_{\odot}). 
\end{equation}
The opacity tables were accordingly chosen based on the AGSS09 metal mixture.
{\small MESA} implements electron conduction opacities \citep{cassisi++2007-opacity} and radiative opacities from OPAL \citep{iglesias+1993-opal,iglesias+1996-opal}, except low-temperature data \citep{ferguson++2005-opacity} and the high-temperature Compton-scattering regime \citep{buchler+1976-opacity}.
The equation of state adopted by {\small MESA} blends from OPAL \citep{rogers+2002-opal-eos}, SCVH \citep{saumon++1995-eos}, PTEH \citep{pols++1995-eos}, HELM \citep{timmes+2000-eos}, and PC \citep{potekhin+2010-eos}.
We adopted nuclear reaction rates from JINA REACLIB database. We only considered a minimal set of elements specified in \texttt{basic.net} of {\small MESA}. 
We did not account for atomic diffusion or gravitational settling in the models. 

For the surface boundary conditions, we used the grey model atmosphere together with Eddington $T-\tau$ integration \citep{eddington-1926-star}.
We caution that by default, {\small MESA} does not include the atmosphere in the output structure. The resulting bias looks very similar to the surface effect, although the amount of correction is larger.
To avoid this, one should specifically set \texttt{add\_atmosphere\_to\_pulse\_data} as \texttt{.true.}

The free parameters for the model grid are 
stellar mass $M\in(0.7,2.3)$ $M_{\odot}$, 
initial helium abundance $Y_{\rm init}\in(0.22, 0.32)$, 
metallicity ${\rm [M/H]}\in(-0.94, 0.56)$ (the corresponding $Z_{\rm init}$ ranges from 0.0016 to 0.0522), 
and mixing-length parameter $\alpha_{\rm MLT}\in(1.3, 2.7)$. 
These four parameters were uniformly sampled in a quasi-random Sobol sequence with a total number of 8191 \citep{bellinger++2016-params-ML}. 
Each set of parameters uniquely determines an evolutionary track. 
Along each evolutionary track, we saved one structure model at least every 0.3 \muHz{} in \Dnu{} or $5$ K in \Teff{}.
For each structure model, we calculated radial mode frequencies with {\small GYRE} in a wide frequency range around $\nu_{\rm max}$.
We used the 6th-order Gauss-Legendre Magnus method to solve the adiabatic oscillations.
We caution that a lower-order algorithm could produce inaccurate frequencies, which differ by an amount larger than the typical observational uncertainties.
Although a higher-order scheme is sensitive to abrupt changes in the structure, we examined the variables (such as density, temperature, sound speed and the first adiabatic index) in the set of oscillation equations, and found they vary smoothly near the atmospheres.


\subsection{Fitting method}\label{subsec:fit-method}
We now describe the fitting method to obtain the surface parameters $\theta_s$ in the prescriptions. 
They determine the amount of surface correction of each model $\theta_m=\{M, Y_{\rm init}, \alpha_{\rm MLT}, {\rm [M/H]}, {\rm age} \}$.
For each star $i$, we considered three classical constraints $q=\{ L, T_{\rm eff}, {\rm [M/H]} \}$ \citep[e.g.][]{valle++2015-ages,joyce+2018-alpha-cen,duanrm++2021-cool-dbv,jiangc+2021-bestp}:
\begin{equation} 
\label{eq:chi2-classical}
\chi^2_{{\rm classical}, i} = \sum_q \frac{\left(q_{{\rm mod}, i}  - q_{{\rm obs}, i}\right)^2 }{\sigma^2_{q, i} }.
\end{equation}
The seismic constraints include radial mode frequencies. They are normalised by the number of observed modes $N_i$, in order to avoid unrealistically small error bars \citep{cunha++2021-modelling-errors,aguirre++2022-basta}:
\begin{equation} 
\chi^2_{{\rm seismic}, i} = \frac{1}{N_i} \sum_n^{N_i} \frac{ \left(\nu_{{\rm mod}, n, i} +\delta\nu_{n,i} -\nu_{{\rm obs},n, i} \right)^2 }{\sigma^2_{{\rm mod}} + \sigma^2_{{\rm obs}, n, i}}.
\end{equation}
Normalising by the number of modes $N_i$ in $\chi^2_{\rm seismic,i}$ is equivalent to reducing the relative weight of $\chi^2_{\rm seismic,i}$ with respect to $\chi^2_{\rm classical,i}$ and artificially inflating returned formal uncertainties. \citet{cunha++2021-modelling-errors} noted that this is a common practice in stellar modelling, but it is not statistically valid and is sometimes unable to capture the systematic uncertainties originating from stellar physics.
In the above equation, $\delta\nu_{n,i}$ is the amount of surface correction, and $\sigma_{{\rm mod}}$ is the systematic uncertainty of stellar model frequencies \citep{litd++2020-kepler-36-subgiants,ong++2021-surf-corr-sg}. 
To evaluate $\sigma_{{\rm mod}}$, we identified the best-fitting model (using the above $\chi^2_{\rm seismic}$ and treating $\sigma_{{{\rm mod}}, i}$ as 0) and calculated its root-mean-square difference between the observed and corrected modelled frequencies. At this step, the amount of surface correction for each mode, $\delta\nu_{n,i}$, was determined by fitting Eq.~\ref{eq:inv-cubic} or~\ref{eq:cubic} to the actual differences between the uncorrected model frequencies $\nu_{{\rm mod},n,i}$ and the observed frequencies $\nu_{{\rm obs},n,i}$ (i.e. the traditional star-by-star surface correction). We then fitted the root-mean-square differences as a function of \numax{} and \Teff{} for the whole sample and used this function to describe $\sigma_{{\rm mod}}$, which gave
\begin{equation}
    \sigma_{{\rm mod}} /\muHz{}= 1.65\cdot (\numax/\numax_{,\odot})^{1.45} (\Teff/\Teff_{,\odot})^{2.30}.
\end{equation}
The value of $\sigma_{{\rm mod}}$ is generally smaller than $\sigma_{{\rm obs}}$ in RGB stars and comparable in main-sequence stars, hence the poorly- and well-observed modes in one star are not weighted similarly.
For our final fitting, $\delta\nu_{n,i}$ was calculated using the prescription described in Sec.~\ref{sec:prescription}.

To obtain the probability distributions of the surface parameters $\theta_s$, we marginalised the probability over other model parameters:
\begin{equation}
    p_i(\theta_s) = \int \exp\left[-\left(\chi^2_{{\rm classical},i} + \chi^2_{{\rm seismic},i} \right)/2\right] \ {\mathrm{d}} \theta_m .
\end{equation}
Since $\theta_m$ is sampled on a pre-computed model grid, in practice, we approximated this integration by taking the average values of the integrated function for all eligible models. 
Finally, putting them together, we maximised the joint probability from all stars in the sample:
\begin{equation} 
 p(\theta_s) = \prod_i  p_i (\theta_s)
\end{equation}
We used a gradient descent algorithm written with {\small JAX} and {\small OPJAX} \citep{jax} to optimise this function. We adopted the uncertainties for the fitted parameters using the diagonal elements of the covariance matrix, which was constructed with the Hessian matrix for $\log p$.
Ensemble modelling to constrain uncertain stellar physics has been used to study the mixing length and helium abundance \citep{lyttle++2021-amlt-he}.

\subsection{Fitting results}\label{subsec:fit-results}
\begin{table*} 
\caption{Best-fitting parameters in the surface correction prescriptions. The stellar models are calculated with $T-\tau$ integrated model atmospheres using the Eddington relation.
\label{tab:surf-corr} \\ 
} 
\addtolength{\tabcolsep}{-2pt}
\begin{tabular*}{1.01\textwidth}{@{\extracolsep{\fill}}lllrrrrrrrrrr}
\toprule
Atmosphere &         Model &  Sample &              $a$ &             $b$ &               $c$ &              $d$ &             $a'$ &             $b'$ &             $c'$ &             $d'$ \\
\midrule
 Eddington &         Cubic &     All & $-4.15 \pm 0.13$ & $0.95 \pm 0.01$ &  $-5.48 \pm 0.23$ & $-1.10 \pm 0.03$ &              --- &              --- &              --- &              --- \\
 Eddington &         Cubic & Pre-RGB & $-4.19 \pm 0.39$ & $0.78 \pm 0.07$ &  $-5.71 \pm 0.58$ & $-0.07 \pm 0.13$ &              --- &              --- &              --- &              --- \\
 Eddington &         Cubic &     RGB & $-4.18 \pm 0.13$ & $0.96 \pm 0.01$ &  $-5.64 \pm 0.22$ & $-1.11 \pm 0.02$ &              --- &              --- &              --- &              --- \\
 Eddington & Inverse-cubic &     All & $-3.74 \pm 0.13$ & $1.09 \pm 0.01$ &  $-8.74 \pm 0.23$ & $-1.38 \pm 0.02$ & $-4.99 \pm 0.14$ &  $1.05 \pm 0.01$ & $-8.15 \pm 0.18$ & $-1.32 \pm 0.02$ \\
 Eddington & Inverse-cubic & Pre-RGB & $-3.72 \pm 0.40$ & $0.61 \pm 0.08$ &  $-1.68 \pm 0.82$ & $-0.08 \pm 0.14$ & $-5.55 \pm 0.42$ &  $0.65 \pm 0.06$ & $-1.57 \pm 0.62$ & $-0.65 \pm 0.11$ \\
 Eddington & Inverse-cubic &     RGB & $-3.84 \pm 0.16$ & $1.10 \pm 0.01$ &  $-8.83 \pm 0.24$ & $-1.38 \pm 0.02$ & $-5.04 \pm 0.17$ &  $1.07 \pm 0.01$ & $-8.33 \pm 0.19$ & $-1.33 \pm 0.02$ \\
\bottomrule
\end{tabular*}
\end{table*}

\begin{figure}
\includegraphics[width=0.5\textwidth]{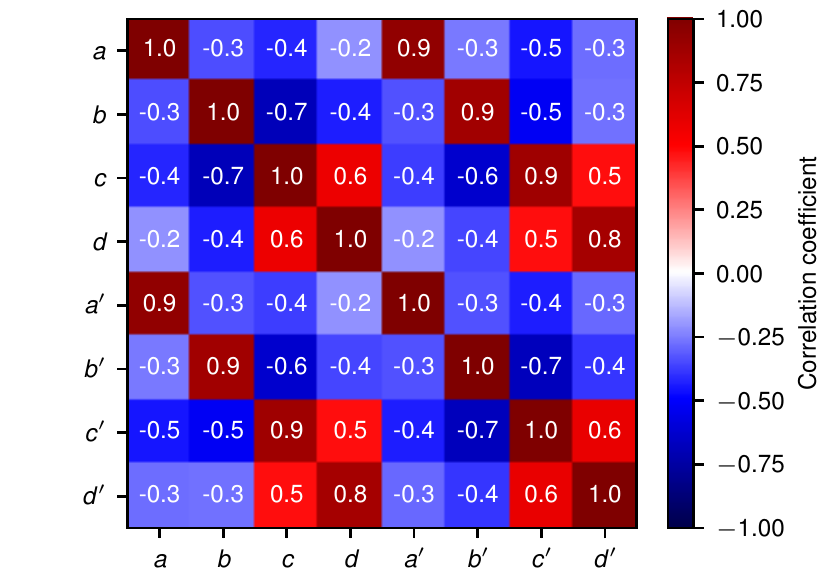}
\caption{Correlation matrix of the fitted parameters for the full sample using the inverse-cubic model.}
\label{fig:corr}
\end{figure}

\begin{figure*}
\includegraphics[width=\textwidth]{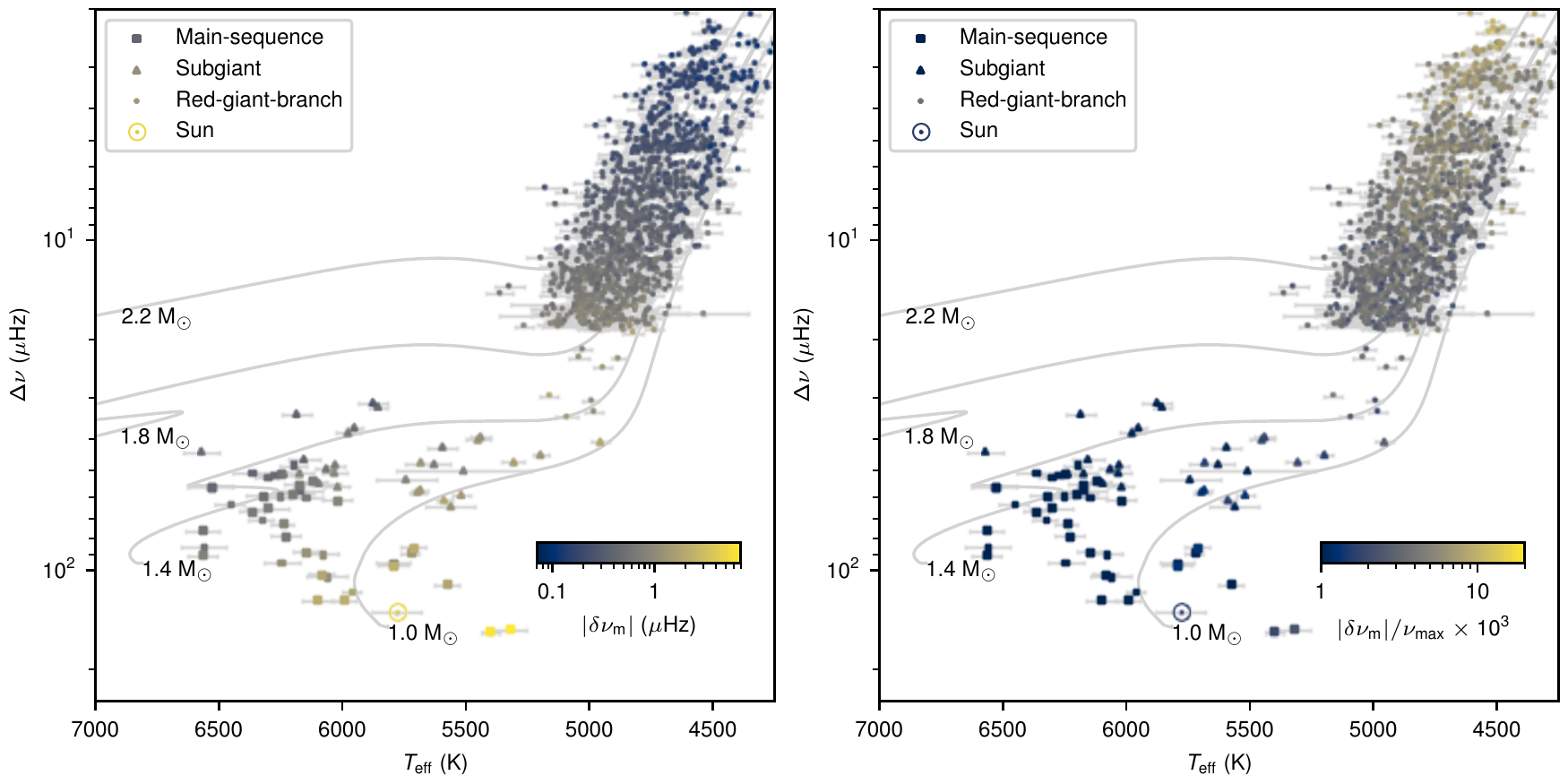}
\caption{Surface correction at \numax{}, \dnum{}, determined based on the prescriptions with the inverse-cubic model, shown on the \Dnu{}--\Teff{} diagrams. The left panel colour-codes the absolute value of \dnum{} (which is negative). The right panel colour-codes the dimensionless quantity, $\dnum/\numax$. The $M=1.0,\ 1.4,\ 1.8$\ and\ $2.2$~\Msolar{} evolutionary tracks are shown in grey lines.}
\label{fig:surf-corr-HR}
\end{figure*}

In addition to fitting the whole sample, we performed fits in two classes of stars: pre-RGB (\numax{}$>$283 \muHz{}) and RGB (\numax{}$<$283 \muHz{}). 
In Table~\ref{tab:surf-corr}, we show the best-fitting values of the surface parameters in the prescriptions. 
Firstly, the best-fitting parameters for the whole sample and the RGB sample are similar, since RGB stars dominate the sample. 
Secondly, the power indices for $g$, \Teff{}, and [M/H] ($b$, $c$, and $d$) are quite different for the RGB fit compared to the pre-RGB fit. 
These parameters are also highly correlated, indicating that their values could be poorly constrained, rather than highly physically different. Fig.~\ref{fig:corr} shows the correlation matrix of the fitted parameters for the full sample using the inverse-cubic model.
Thirdly, although the reported uncertainties are small, we observed strong correlations (correlation coefficient greater than 0.5) between $b$ and $c$, $b'$ and $c'$, $c$ and $d$, $c'$ and $d'$, $a$ and $a'$, $b$ and $b'$, $c$ and $c'$, and $d$ and $d'$. 
Fourthly, the inverse-cubic and the cubic models show little differences.  
We discuss the inverse-cubic model and the fit with the whole sample in the rest of the paper.

To visualise our fitting result, we colour-coded the values of \dnum{} and $\dnum/\numax$ in Fig.~\ref{fig:surf-corr-HR} on the \Dnu{}--\Teff{} diagrams.
In terms of the absolute value \dnum{} (which is always negative), the main-sequence stars have the largest amount of surface correction. 
It decreases towards higher \Teff{} (hotter F-stars) and smaller \Dnu{} (more luminous red giants).
Concerning the relative value of \dnum{} with respect to \numax{}, the trend is reversed.
The main-sequences stars have smaller corrections, and the surface effect becomes increasingly significant for luminous red giants.
Those trends are similar compared to those found by \citet[][Fig.~5 and~6]{trampedach++2017-surface-effect-3d-simulation-1-convective-expansion-atmos}, who improved the mean atmospheric structure with 3D-averaged models (the so-called ``structural effect'').

\citet[][Eqs. 9 and 10]{sonoi++2015-surf-corr-3d} also studied the structural effect, mainly for dwarfs and subgiants, and concluded positive correlations between \dnum{} and $g$ or \Teff{}. \citet[][Fig. 5]{houdek++2019-damping-rate-lagacy-kepler} studied the ``modal effect'', which accounts for the coupling between convection and oscillation, and reported a similar correlation with their 1D time-dependent convection models.
These works are qualitatively consistent with our best-fitting parameters for dwarfs and subgiants (Table~\ref{tab:surf-corr}).

We emphasise that the values reported in Table~\ref{tab:surf-corr} may not be directly applicable to other stellar models, which could have different outer boundary conditions. 
For example, in Fig. 2 of \citet{jcd++1996-current-state-solar-modelling-science} there are three modifications to the solar atmosphere: one with an alternative treatment of the convective flux, one with the inclusion of turbulent pressure, and one with replacement from 3D averaged models. 
Each one has a different value for the surface effect at \numax{}, ranging from 5 to 17 \muHz{}, suggesting that an alternative atmosphere could differ by a factor of three from our fitted values for the Sun.
Moreover, the differences in the model physics and even numerical treatment can change the fitted values. 
To test this, we applied our prescriptions to a grid of models calculated by \citet{sharma++2016-population-rg-kepler}. Even though they set the same Eddington atmosphere with MESA as in this work, we obtained unrealistically large corrections in their models at high radial orders. 
Additionally, Appendix~\ref{sec:hopf} examines an alternative atmosphere based on the Hopf $T-\tau$ relation. We found that the fitted coefficients are drastically different from those obtained using the Eddington relation.
Hence, to correctly implement our method, we recommend to either use the stellar models corrected in this work or to re-fit the prescriptions with stellar models of the user's choice. 

\section{Improvements on parameter estimations}
\label{sec:params}

\begin{figure}
\includegraphics[width=\columnwidth]{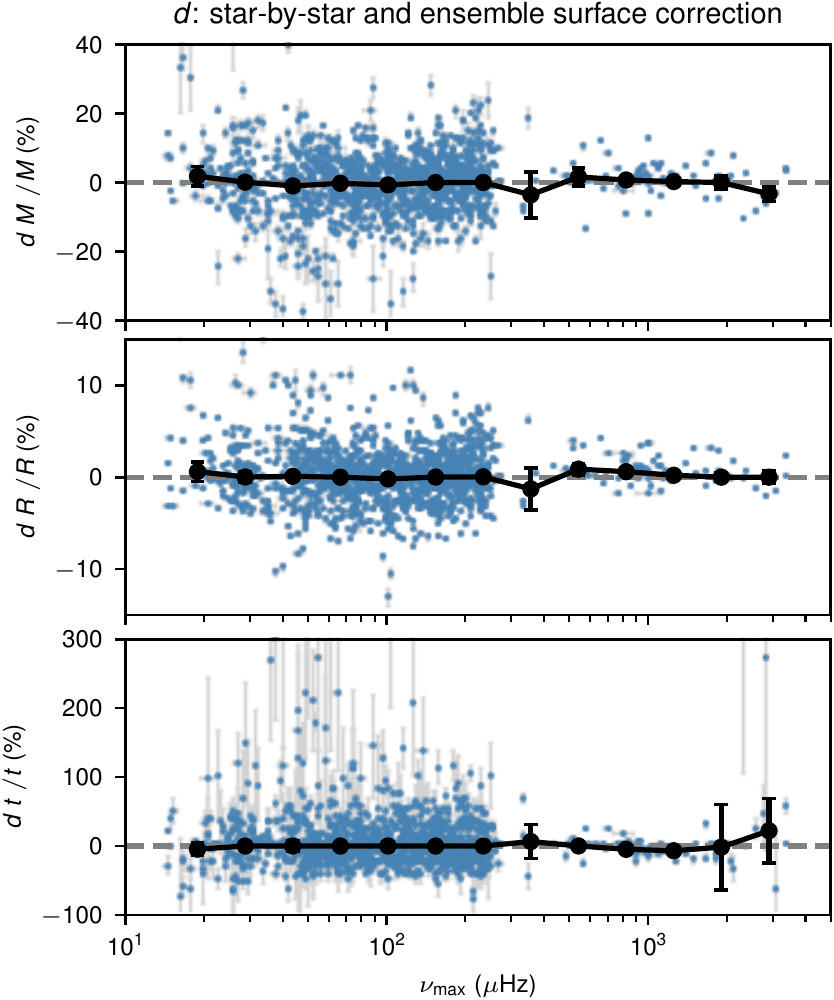}
\caption{Fractional differences of mass, radius, and age, between modelling using star-by-star and ensemble surface correction. The medians and the associated error bars in bins of equal width in $\log{\numax}$ are shown in black circles.}
\label{fig:params}
\end{figure}

\begin{figure}
\includegraphics[width=\columnwidth]{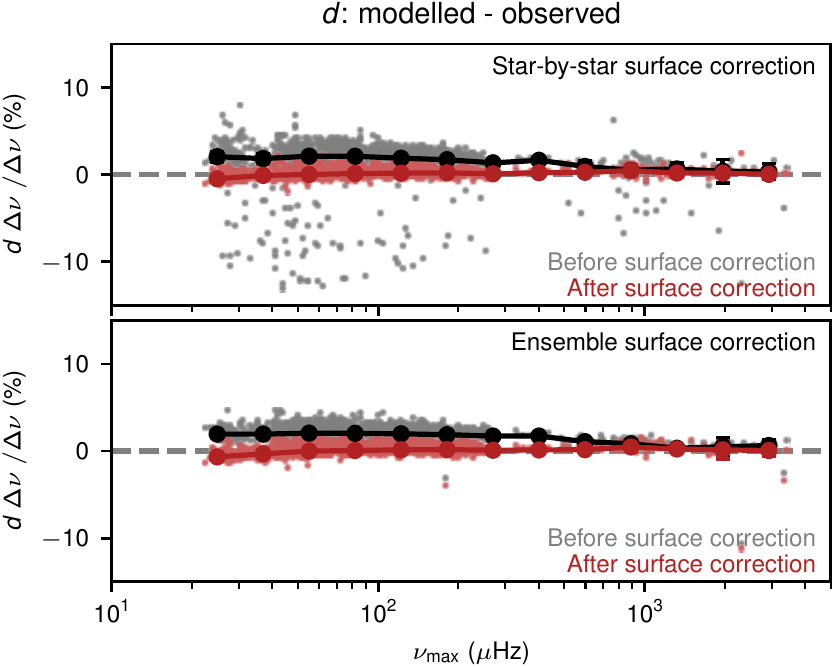}
\caption{Fractional differences of theoretical and observational \Dnu{}. The top panel shows results from star-by-star fits. The bottom panel shows those from ensemble fits.}
\label{fig:Dnu-compare}
\end{figure}

\begin{figure}
\includegraphics[width=\columnwidth]{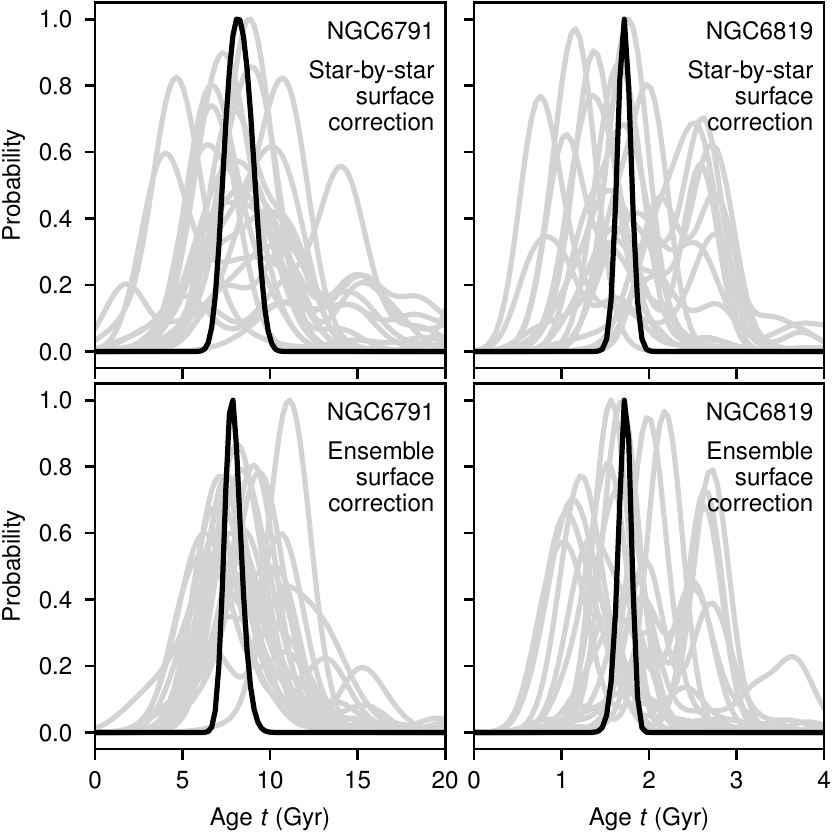}
\caption{Probability distributions of stellar ages for the RGB stars within the \kepler{} open clusters NGC 6791 and NGC 6819. The age probability for each star are shown in grey, while the joint probability distribution is shown in black. Modelling with the prescription (lower panels) shows a reduction of the scatter in age, compared to modelling with a star-by-star surface correction (upper panels).}
\label{fig:clusters}
\end{figure}

We now check whether applying our prescription introduces bias in the estimated stellar properties. 
In Fig.~\ref{fig:params}, we show the fractional differences of mass, radius, and age, between modelling without and with the prescription. 
The differences have medians fluctuating around 0, suggesting no systematic bias. 

Next, we demonstrate two major improvements by using our method.
Firstly, we note that adopting the prescriptions reduces outliers when inferring parameters from stellar modelling.
For example, Fig.~\ref{fig:params} shows some stars significantly away from the median values. 
These data points correspond to a poor fit due to the unconstrained surface correction. 
To confirm this, we show the differences between the modelled and observed values of \Dnu{} in Fig.~\ref{fig:Dnu-compare}, where the former were obtained from the best-fitting model.
The model \Dnu{} values were determined from the slope of a linear fit to the radial frequencies versus the orders, with weights of each mode assigned by a Gaussian envelope (centred around \numax{}; see \citealt{white++2011-asteroseismic-diagrams-cd-epsilon-deltaP-models})
\begin{equation}
    w = \exp\left[-\frac{(\nu-\numax)^2}{2 \sigma^2}\right],
\end{equation}
where $\ln\sigma/\muHz = 0.964 \ln\numax/\muHz - 1.715$, which was estimated based on a fit to observations \citep{yuj++2018-16000-rg,lund++2017-legacy-kepler-1,liyg++2020-kepler-36-subgiants}. 
The values are similar to those obtained by \citet{mosser++2012-amplitude-rg}. 
For all models in this work, we calculated the modes within $5\sigma$ of \numax{}.

By comparing the red points in the two panels, we noticed that the differences are similar after the surface correction, independent of whether using the prescription or not.
This is expected, since the corrected frequencies were constructed to fit with the observed frequencies.
However, when we compare the grey points in both panels, which represent \Dnu{} calculated from the uncorrected frequencies, the outliers are only present in the case of star-by-star fit (top panel).
These outliers are eliminated when the prescription was applied in the ensemble fit (bottom panel).

Secondly, we argue that adopting the prescriptions also reduces scatter in model-based parameters. 
This can be seen from modelling stars in open clusters, members of which are expected to have the same age.
We examined the test sample introduced in Sec.~\ref{subsec:obs}, namely the RGB stars in NGC 6791 and NGC 6819.
In Fig.~\ref{fig:clusters}, we show the probability distributions of ages for each star (in grey lines), and compare them with and without using our new prescription. 
It is evident that the probabilities with the prescriptions applied display smaller scatter overall.
The root-mean-square values of individual model-based ages reduces from 2.11 Gyr to 1.60 Gyr for NGC 6791, and from 0.46 Gyr to 0.34 Gyr for NGC 6819.

\begin{table} 
\caption{List of ages for the two open clusters. \label{tab:cluster-ages} }
\begin{tabular*}{0.48\textwidth}{@{\extracolsep{\fill}}lcr}
\toprule

Age (Gyr)   & Methods  & References \\

\midrule
\multicolumn{3}{c}{NGC 6791}      \\
\midrule
$6.8$ -- $8.6$  & Seismology & \citet{basu++2011-model-ngc6791-ngc6819} \\
$7.68 \pm 1.60$ & Seismology & This work \\
$8.2 \pm 0.3$ & Seismology & \citet{mckkever++2019-ngc6791} \\
$8.3 \pm 0.8$ & Isochrone fitting & \citet{brogaard++2012-ngc6791}\\
$8.3$ & Binary & \citet{brogaard++2021-ngc6791} \\
$10.1 \pm 0.9$ & Seismology & \citet{kallinger++2018-sc} \\
\midrule 
\multicolumn{3}{c}{NGC 6819} \\
\midrule 
$1.57\pm0.34$ & Seismology &This work \\
$2$ -- $2.4$ & Seismology &\citet{basu++2011-model-ngc6791-ngc6819} \\
$2.4$ & Isochrone fitting & \citet{jeffries++2013-ngc6819} \\
$2.4\pm0.2$ & Eclipsing binary & \citet{brewer++2016-ngc6819} \\
$2.5$ & Isochrone fitting & \citet{balona++2013-ngc6819} \\
$2.62 \pm 0.25$ & Eclipsing binary & \citet{sandquist++2013-ngc6819} \\
$2.9 \pm 0.3$ & Seismology &\citet{kallinger++2018-sc} \\
$3.1\pm0.4$ & Eclipsing binary& \citet{jeffries++2013-ngc6819} \\
\bottomrule
\end{tabular*}
\end{table}

On a side note, we compared our age estimates for the two open clusters against recent studies in Table~\ref{tab:cluster-ages}. 
Our estimates occupy the lower end of the published ages. 
We attribute this to our adopted observational constraints: the relatively high metallicity for NGC 6791 ([M/H] $=0.36$ dex), and the large extinction value for NGC 6819 (mean $E(B-V) =0.17$). The effect of extinction and metallicity on age can be seen from Table 2 of \citet{basu++2011-model-ngc6791-ngc6819}.

To allow easy-access to our stellar models and the surface-corrected frequencies, we published them in an online repository (see Data Availability). 

\section{Correction to the p-mode large separation from stellar models}
\label{sec:dnu}

\begin{figure}
\includegraphics[width=\columnwidth]{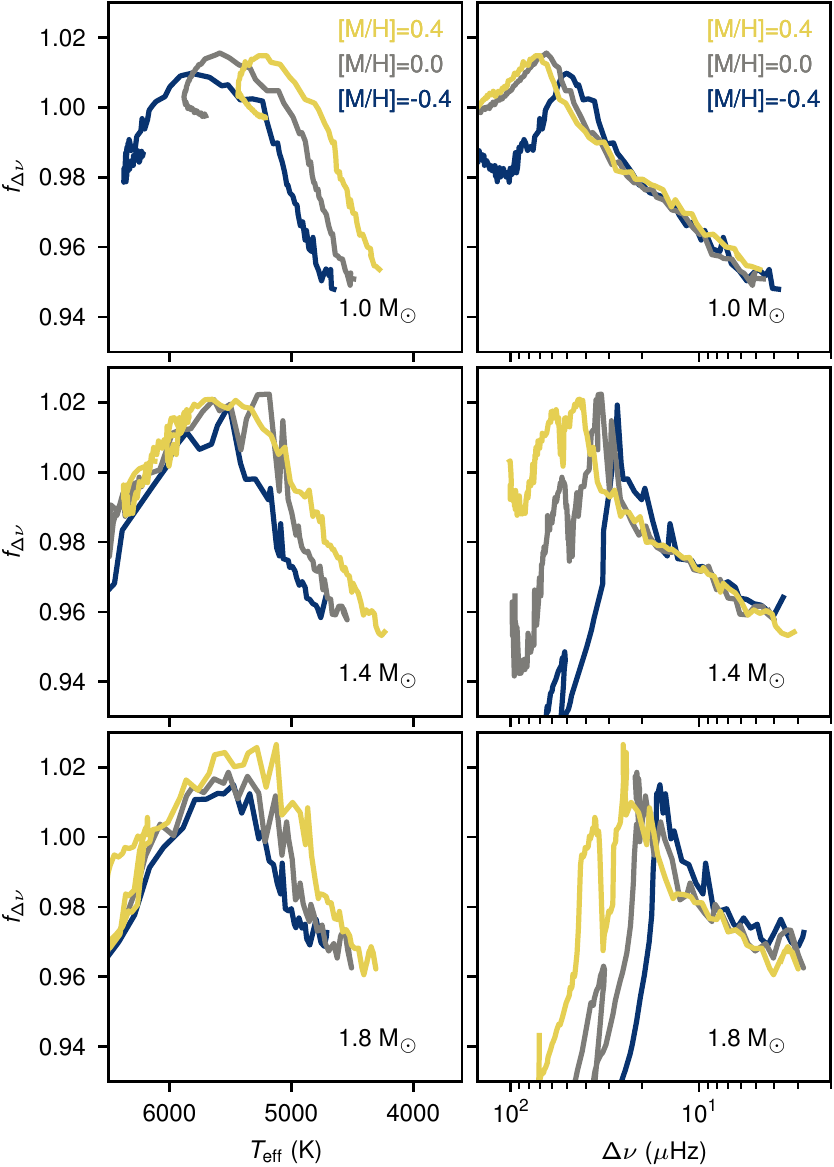}
\caption{Correction factor for the \Dnu{} scaling relation, \fDnu{}, as a function of \Teff{} (left panels) and \Dnu{} (right panels) for three metallicities and three masses. The values for \fDnu{} were derived using stellar models with the surface correction considered. The small fluctuations along the lines arise from the uncertainty in the helium abundance and the mixing length parameter. }
\label{fig:fDnu}
\end{figure}

\begin{figure}
\includegraphics[width=\columnwidth]{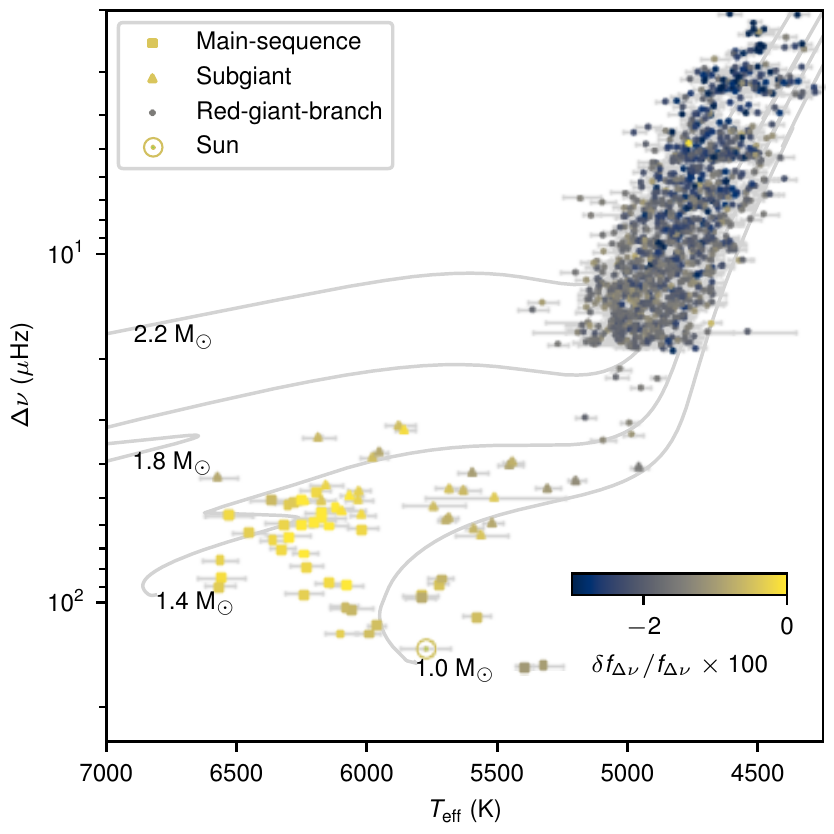}
\caption{Fractional differences of \fDnu{} between before and after the surface correction. }
\label{fig:d_fDnu_HRD}
\end{figure}

\begin{figure}
\includegraphics[width=\columnwidth]{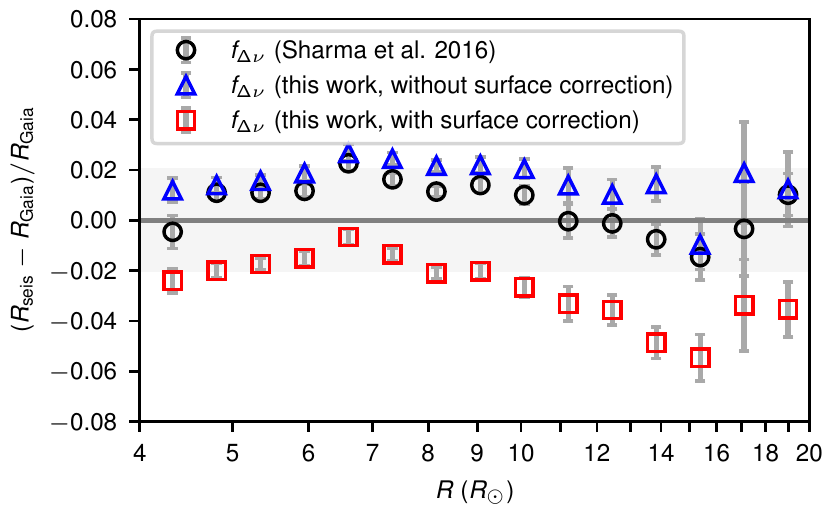}
\caption{Comparison of \Gaia{} radii and asteroseismic radii, using \fDnu{} calculated in \citetalias{sharma++2016-population-rg-kepler} and this work. The data points are binned medians, and the error bars represent the standard errors of the medians. The grey band highlights the $2\%$ systematic uncertainties (e.g. temperature scale) discussed by \citet{zinn++2019-radius-sc}.}
\label{fig:gaia}
\end{figure}

\begin{figure}
\includegraphics[width=\columnwidth]{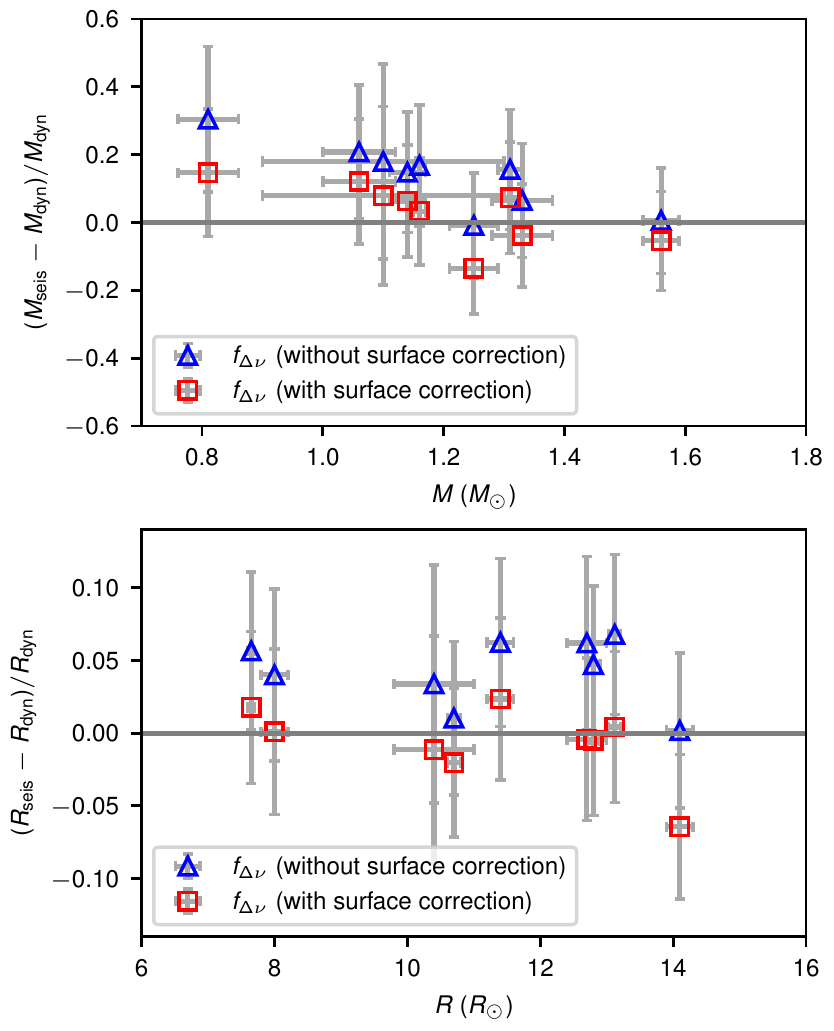}
\caption{Comparison of the dynamical and the asteroseismic masses and radii using eclipsing binaries \citep{gaulme++2016-eb-sc,themebl++2018-rg-3ebs,brogaard++2018-accuracy-scaling-relation,benbakoura++2021-binary}. The asteroseismic properties were determined using \fDnu{} with and without the surface correction, respectively.}
\label{fig:ebs}
\end{figure}

The scaling relation that relates the p-mode large separation \Dnu{} to stellar mean density, $\Dnu\propto\sqrt{\rho}$ \citep{ulrich-1986-age}, is broadly used \citep[see][for a review]{hekker-2020-scaling-review}. This relation is only an approximation and stellar models have been used to correct it \citep{white++2011-asteroseismic-diagrams-cd-epsilon-deltaP-models,sharma++2016-population-rg-kepler,guggenberger++2016-metallicity-scaling-relation,rodrigues++2017-dpi-modelling,serenelli++2017-apokasc-dwarf-subgiant,pinsonneault++2018-apokasc}. \citet[][hereafter S16]{sharma++2016-population-rg-kepler} introduced a correction factor \fDnu{} to the standard \Dnu{} scaling relation:
\begin{equation}\label{eq:Dnu}
    \left( \frac{\Dnu}{\Dnu_\odot}\right) = \fDnu \left(\frac{\rho}{\rho_{\odot}}\right)^{0.5},
\end{equation}
where $\Dnu_\odot = 135.1$~\muHz{} is the solar value of the large frequency separation \citep{huber++2011-scaling-kepler}.

The correction factor \fDnu{} are used when estimating the mass and radius via the usual scaling relations \citep{stello++2008-wire-kgiants,kallinger++2010-rg-corot-mass-radius}:
\begin{equation}
\label{eq:sc-mass}
    \frac{M}{\Msolar} \approx \left(\frac{\numax}{\numax_{,\odot}}\right)^3 \left(\frac{\Dnu}{\fDnu\Dnu_\odot} \right)^{-4} \left(\frac{\Teff}{\Teff_{,\odot}}\right)^{3/2},
\end{equation}
and
\begin{equation}
\label{eq:sc-radius}
    \frac{R}{\Rsolar} \approx \left(\frac{\numax}{\numax_{,\odot}}\right) \left(\frac{\Dnu}{\fDnu\Dnu_\odot} \right)^{-2} \left(\frac{\Teff}{\Teff_{,\odot}}\right)^{1/2}.
\end{equation}

To use Eq.~\ref{eq:Dnu} to determine \fDnu{} from models, we need to know the model-predicted density and \Dnu{}, the latter of which is usually calculated from radial oscillation frequencies. 
Since the surface correction is negative, we expect that the model \Dnu{} value will decrease when the correction is applied \citep{kjeldsen++2008-surface-correction}. 
However, this correction was previously ignored.
Here, we investigate this change and analyse its implication on stellar properties derived from the asteroseismic relations.

\subsection{Results}\label{subsec:fdnu-results}

Firstly, we present the correction factor, \fDnu{}, calculated from our stellar models and prescriptions. 
Fig.~\ref{fig:fDnu} shows \fDnu{} as a function of \Teff{} (left panels) and \Dnu{} (right panels), for three masses and three metallicities. 
The overall variations of \fDnu{} resemble those calculated from models by \citet[][Fig. 4]{white++2011-asteroseismic-diagrams-cd-epsilon-deltaP-models} and \citetalias{sharma++2016-population-rg-kepler} (Fig. 4), neither of which included a surface correction.
However, our values for \fDnu{} are systematically smaller compared to the work from \citetalias{sharma++2016-population-rg-kepler}, due to this correction. 
Fig.~\ref{fig:d_fDnu_HRD} shows the effect of surface correction on \fDnu{}.
The change of \fDnu{} is small for main-sequence stars, but is larger on the RGB, showing an $\sim$2\% reduction, where the surface correction is relatively significant (see also Fig.~\ref{fig:surf-corr-HR}b). 

The revised correction factors can be used to estimate the mass and radius via Eq.~\ref{eq:sc-mass} and~\ref{eq:sc-radius}.
Unlike the surface corrections done in Sec.~\ref{subsec:fit-method}, these do not require any additional model calculations by the user. 
They simply involve revising the standard scaling relation. 
We provide a {\small Python} routine to derive \fDnu{} given user-specified observables, based on the models that are calibrated in this work.
For a given star, the user specifies observational constraints and their associated uncertainties (e.g. $L$, \Teff{}, [M/H] and \numax{}).
Each model is assigned with a $\chi^2$ (using Eq.~\ref{eq:chi2-classical}). 
Next, the correction factor \fDnu{} of the star is estimated by taking the average of model \fDnu{} values, weighted by $\exp(-\chi^2/2)$.

We can also provide a simple fitting formula of \fDnu{} with respect to stellar properties. We explored various functional forms (linear, log-linear and polynomial) and included the observed \numax{}, \Dnu{}, \Teff{} and [M/H] as independent variables to perform simple regressions. The following form obtains a reasonably good fit ($r^2=0.85$) and avoids over-fitting with higher orders (examined via cross-validation):
\begin{equation}\label{eq:fDnu}
\begin{aligned}
    f_{\Dnu} = \beta_0 &+  \beta_1\log_{10}(\numax/3090\ \muHz) \\
    &+\beta_2\log_{10}(\Dnu/135.1\ \muHz) \\
    &+\beta_3(\Teff/5777{\ \rm K}) \\
    &+\beta_4(\Teff/5777{\ \rm K})^2 \\
    &+\beta_5(\Teff/5777{\ \rm K})^3 \\
    &+\beta_6{\rm [M/H]}, \\ 
    & \text{for}\ 0.8<M/\Msolar<2.2,\ -0.8<{\rm [M/H]}<0.5,  \\
    & \text{and pre-RGB tip}\  (\Dnu>2.0\ \muHz).\\
\end{aligned}
\end{equation}
The best-fitting parameters are $\beta=\{ 
\allowbreak 4.015 \allowbreak\pm 0.225, 
\allowbreak0.168 \allowbreak\pm 0.012, 
\allowbreak-0.186 \allowbreak\pm 0.015, 
\allowbreak-10.234 \allowbreak\pm 0.735, 
\allowbreak11.432 \allowbreak\pm 0.801, 
\allowbreak-4.200 \allowbreak\pm 0.290, 
\allowbreak0.001 \allowbreak\pm 0.001\}$.
Note that the above formula does not predict $\fDnu=1$ for solar properties, since we did not require the formula to pass through the solar reference point.
To obtain the most accurate estimations on \fDnu{}, we suggest using the provided {\small Python} routine.

Naturally, fitting formula would need to be tested before applying outside the parameter ranges of our sample. 
Extending it to other ranges, such as metal-poor, high-mass, and red-clump stars, requires more data and will be the subject of future work.

\subsection{Comparisons to Gaia radii}\label{subsec:gaia}
To show how the surface corrected \fDnu{} affects stellar radii, we compared the asteroseismic radii with the Gaia radii calculated by \citet{zinn-2021-gaia-dr3-plx-seismology}, using the APOKASC sample \citep{pinsonneault++2018-apokasc}. 
The \Teff{} and [M/H] from APOGEE \citep{abdurrouf++2021-apogee-dr17} were used to derive \Gaia{} radii for bolometric corrections and converting from luminosities. 
We calculated the asteroseismic radii through Eq.~\ref{eq:sc-radius}, where we adopted the SYD pipeline values for \Dnu{} and \numax{} \citep{serenelli++2017-apokasc-dwarf-subgiant,yuj++2018-16000-rg}, and \Teff{} from APOGEE.

Fig.~\ref{fig:gaia} shows the result. 
Without the surface correction considered, the \fDnu{} in this work (blue triangles) produce similar radii to \citetalias{sharma++2016-population-rg-kepler} (black circles), despite the fact that the underlying stellar models are different. 
\citet{jcd++2020-aarhus-rgb-osc} reported a spread of only 0.2\% in the values of \fDnu{} from different stellar modelling code.
In addition, we found that differences in the mixing length can change \fDnu{} by $\sim$1\% (see Appendix~\ref{sec:fdnu-error} for more details).
A larger discrepancy emerges when we applied the surface correction (red squares).
Eq.~\ref{eq:sc-radius} indicates that the seismic radius is proportional to $f_{\Delta\nu}^2$. 
Since correctly accounting for the surface effect reduces \fDnu{} by $\sim$2\% for RGB stars (Fig.~\ref{fig:d_fDnu_HRD}), this translates to a systematic $\sim$4\% decrease of the asteroseismic radius scale. 
This is exactly what we see in Fig.~\ref{fig:gaia}. 
As summarized by \citet{zinn++2019-radius-sc}, the systematic uncertainties involved in this comparison, such as uncertainties in bolometric correction and extinction, the IRFM temperature scale, and asteroseismic reference points, can add up to 2\%. 
It is not yet possible to conclude any disagreement between the asteroseismic and \Gaia{} radii with this precision.

There is a significant excursion of red squares at $R>10$~\Rsolar{} in Fig.~\ref{fig:gaia} that cannot be explained by the causes discussed above. 
The dip associated with the excursion is also present when using the \fDnu{} without surface correction. 
The red clump stars, which burn helium in the core, have radii around $10$~\Rsolar{} \citep[see Fig. 6 of][]{liyg++2021-sc-intrinsic-scatter}. 
After exhausting core helium, they become asymptotic-giant-branch (AGB) stars and are difficult to be distinguished from RGB stars based on g-mode period spacings alone \citep{kallinger++2012-epsp,dreau++2021-agb-rgb}. 
Hence, the dip is probably a result of the contamination from AGB stars in the sample. 
This is supported by the excess of very low-mass stars above $10$~\Rsolar{}, due to AGB stars having lost more mass than RGB stars \citep[see Fig. 2g of][]{liyg++2021-sc-intrinsic-scatter}.
The impact of this contamination on Galactic population studies clearly deserves further investigation.

\subsection{Comparisons to eclipsing binaries}\label{subsec:ebs}

Similarly, we can compare the asteroseismic radius and mass with the dynamical properties determined from eclipsing binaries. 
We used the eclipsing binary sample studied by \citet{gaulme++2016-eb-sc}, \citet{themebl++2018-rg-3ebs}, \citet{brogaard++2018-accuracy-scaling-relation} and \citet{benbakoura++2021-binary}, who determined the dynamical masses and radii through radial-velocity and lightcurve modelling.
We calculated their asteroseismic radii and masses through Eq.~\ref{eq:sc-mass} and~\ref{eq:sc-radius}, using \Dnu{}, \numax{}, and \Teff{} reported in \citet{benbakoura++2021-binary}.
Fig.~\ref{fig:ebs} shows the resulting comparison. 
Using the corrected \fDnu{} produces excellent agreement of those properties determined from the two independent means, while the \fDnu{} without correction tends to systematically overestimate them.
Although \citet{benbakoura++2021-binary} did not consider the surface corrected \fDnu{}, they also found agreement between the asteroseismic and dynamical properties, through modifying the solar reference values appearing in the scaling relations. 
Our results thus remove the need to shift the reference values when the surface correction is taken into account.
Moreover, the impact of surface correction changes as a function of stellar properties (Fig.~\ref{fig:d_fDnu_HRD}), so it would be difficult to reconcile all stars if the reference values are treated as a constant.

According to Eq.~\ref{eq:sc-mass}, the scaling mass is proportional to $f_{\Dnu}^4$\citep[e.g.][]{sharma++2016-population-rg-kepler}. 
Hence, as a result of the change in \fDnu{}, the seismic mass scale decreases by $\sim$8\%. 
This could have significant consequences for Galactic archaeology since the ages of low-mass stars are critically dependent on their masses.

\section{Conclusions}
\label{sec:conc}

We provide a prescription for the surface correction as a function of stellar properties, exploiting the fact that the correction should vary smoothly across the H--R diagram. 
Our main findings are summarised as follows:
\begin{enumerate}
    \item The absolute values of the surface correction are larger in main-sequence stars and smaller in RGB stars. For the relative surface correction as a fraction of \numax{}, the trend is reversed (Sec.~\ref{subsec:fit-results} and Fig.~\ref{fig:surf-corr-HR}). 
    \item Using the prescription, we were able to reduce scatter and the number of outliers in stellar properties estimated from stellar modelling (Sec.~\ref{sec:params} and Figs.~\ref{fig:Dnu-compare} and~\ref{fig:clusters}). This demonstrates the power of our ensemble-based parameterization of the surface correction.
    \item We present our stellar models in an online repository. The models include radial frequencies before and after applying the surface correction calibrated in this work.
    \item Taking into account the surface correction, we present a revised \Dnu{} scaling relation (Sec.~\ref{subsec:fdnu-results} and Fig.~\ref{fig:fDnu}). We provided a fitting formula (Eq.~\ref{eq:fDnu}) and a {\small Python} routine to determine \fDnu{} given user-specified observables.
    \item The values of \fDnu{} are smaller by up to 2\%, after taking into account the surface correction (Sec.~\ref{subsec:fdnu-results} and Fig.~\ref{fig:d_fDnu_HRD}). This results in decreases of up to 4\% in radii and up to 8\% in masses when using the asteroseismic scaling relations. 
    \item We showed that the mass and radius determined with the revised \fDnu{} improve the agreement with those determined from eclipsing binaries (Sec.~\ref{subsec:ebs} and Fig.~\ref{fig:ebs}).
\end{enumerate}
For most readers, item (iv) will be the most useful. It describes a modification to the \Dnu{} scaling relation that, for the first time, takes the surface correction into account and we encourage its use when deriving masses and radii from asteroseismic parameters.

\section*{Acknowledgements}
We thank Warrick Ball, Sarbani Basu, Karsten Brogaard, Jørgen Christensen-Dalsgaard, and G\"{u}nter Houdek for interesting comments and discussions.
T.R.B acknowledges funding from the Australian Research Council (Discovery Project DP210103119 and Laureate Fellowship FL220100117). D.H. acknowledges support from the Alfred P. Sloan Foundation and the National Aeronautics and Space Administration (80NSSC19K0597). M.J. acknowledges the Lasker Fellowship granted by the Space Telescope Science Institute. T.L. acknowledges the Joint Research Fund in Astronomy (U2031203) under cooperative agreement between the National Natural Science Foundation of China (NSFC) and Chinese Academy of Sciences (CAS) and the NSFC grants 12090040 and 12090042. 
H.K. acknowledges funding for the Stellar Astrophysics Centre provided by The Danish National Research Foundation (Grant agreement no.: DNRF106).

We gratefully acknowledge the Kepler teams, whose efforts made these results possible.
Funding for the Kepler mission is provided by the NASA Science Mission Directorate. This paper includes data collected by the Kepler mission and obtained from the MAST data archive at the Space Telescope Science Institute (STScI). STScI is operated by the Association of Universities for Research in Astronomy, Inc., under NASA contract NAS 5–26555.

This work presents results from the European Space Agency (ESA) space mission Gaia. Gaia data are being processed by the Gaia Data Processing and Analysis Consortium (DPAC). Funding for the DPAC is provided by national institutions, in particular the institutions participating in the Gaia MultiLateral Agreement (MLA). The Gaia mission website is https://www.cosmos.esa.int/gaia. The Gaia archive website is https://archives.esac.esa.int/gaia.

Funding for the Sloan Digital Sky Survey IV has been provided by the Alfred P. Sloan Foundation, the U.S. Department of Energy Office of Science, and the Participating Institutions. 

We acknowledge the use of the National Computational Infrastructure (NCI) which is supported by the Australian Government, and accessed through the Sydney Informatics Hub HPC Allocation Scheme, which is supported by the Deputy Vice-Chancellor (Research), University of Sydney. 


This work is made possible by the following open-source software: {\small Numpy} \citep{numpy}, {\small Scipy} \citep{scipy}, {\small Matplotlib} \citep{matplotlib}, {\small Astropy} \citep{astropy1,astropy2}, {\small Pandas} \citep{pandas}, {\small MESA} \citep{paxton++2011-mesa,paxton++2013-mesa,paxton++2015-mesa,paxton++2018-mesa,paxton++2019-mesa}, {\small GYRE} \citep{townsend+2013-gyre}, {\small ISOCLASSIFY} \citep{huber++2017-seismic-radii-gaia,berger++2020-gaia-kepler-1-stars}, {\small JAX} and {\small OPTAX} \citep{jax}, {\small tqdm} \citep{tqdm}, \verb|tensorflow|. 

\section*{Data Availability}
The code and processed data used in this work are available on Github.\footnote{https://www.github.com/parallelpro/surface}
The calibrated stellar models including oscillation frequencies and the correction factors \fDnu{} can be downloaded from zenodo.\footnote{https://zenodo.org/record/7905521}
All raw data (e.g. the Keck spectra) are available on request to the corresponding authors.




\bibliographystyle{mnras-nl}
\bibliography{references/myastrobib.bib} 

\appendix

\section{Surface correction with Hopf atmospheric models}
\label{sec:hopf}

We carried out an additional set of stellar model calculation using the solar-calibrated Hopf atmosphere \citep{paxton++2013-mesa}, which is equivalent to the fit provided by \citet{sonoi++2019-mlt} to the VAL-C model \citep{vernazza++1981-valc}. The results are presented in Table~\ref{tab:surf-corr-hopf}. 
Notably, there is a significant discrepancy in the fitted coefficients between the pre-RGB and RGB samples --- especially for $a$, which represents the amount of surface correction at \numax{} for a solar model. 
Considering the atmosphere is solar-calibrated, it may generalise poorly on RGB stars. 
Further studies are needed to understand its cause.
To enhance the applicability of the surface correction prescription proposed in this paper, we recommend using the atmosphere calculated based on the Eddington $T-\tau$ relation rather the Hopf atmosphere.

\begin{table*} 
\caption{Best-fitting parameters in the surface correction prescriptions. The stellar models are calculated with $T-\tau$ integrated model atmospheres using the solar-calibrated Hopf relation.
\label{tab:surf-corr-hopf} \\ 
} 
\addtolength{\tabcolsep}{-3pt}
\begin{tabular*}{1.01\textwidth}{@{\extracolsep{\fill}}lllrrrrrrrrrr}
\toprule
Atmosphere &         Model &  Sample &              $a$ &             $b$ &               $c$ &              $d$ &             $a'$ &             $b'$ &             $c'$ &             $d'$ \\
\midrule
      Hopf &         Cubic &     All & $-3.06 \pm 0.10$ & $0.96 \pm 0.01$ &  $-7.14 \pm 0.20$ & $-0.88 \pm 0.02$ &              --- &              --- &              --- &              --- \\
      Hopf &         Cubic & Pre-RGB & $-4.25 \pm 0.28$ & $0.50 \pm 0.06$ &  $-0.98 \pm 0.49$ & $-0.49 \pm 0.08$ &              --- &              --- &              --- &              --- \\
      Hopf &         Cubic &     RGB & $-2.90 \pm 0.10$ & $0.97 \pm 0.01$ &  $-7.68 \pm 0.22$ & $-0.91 \pm 0.02$ &              --- &              --- &              --- &              --- \\
      Hopf & Inverse-cubic &     All & $-4.73 \pm 0.40$ & $0.45 \pm 0.06$ &   $2.27 \pm 0.43$ &  $0.66 \pm 0.11$ & $-6.01 \pm 0.38$ & $-6.01 \pm 0.38$ & $-6.01 \pm 0.38$ & $-6.01 \pm 0.38$ \\
      Hopf & Inverse-cubic & Pre-RGB & $-4.73 \pm 0.40$ & $0.45 \pm 0.06$ &   $2.27 \pm 0.43$ &  $0.66 \pm 0.11$ & $-6.01 \pm 0.38$ & $-6.01 \pm 0.38$ & $-6.01 \pm 0.38$ & $-6.01 \pm 0.38$ \\
      Hopf & Inverse-cubic &     RGB & $-1.45 \pm 0.03$ & $0.88 \pm 0.01$ & $-10.16 \pm 0.20$ & $-0.31 \pm 0.02$ & $-2.31 \pm 0.06$ & $-2.31 \pm 0.06$ & $-2.31 \pm 0.06$ & $-2.31 \pm 0.06$ \\
\bottomrule
\end{tabular*}
\end{table*}

\section{Systematic uncertainties of the correction factors}\label{sec:fdnu-error}

\begin{figure}
\includegraphics[width=\columnwidth]{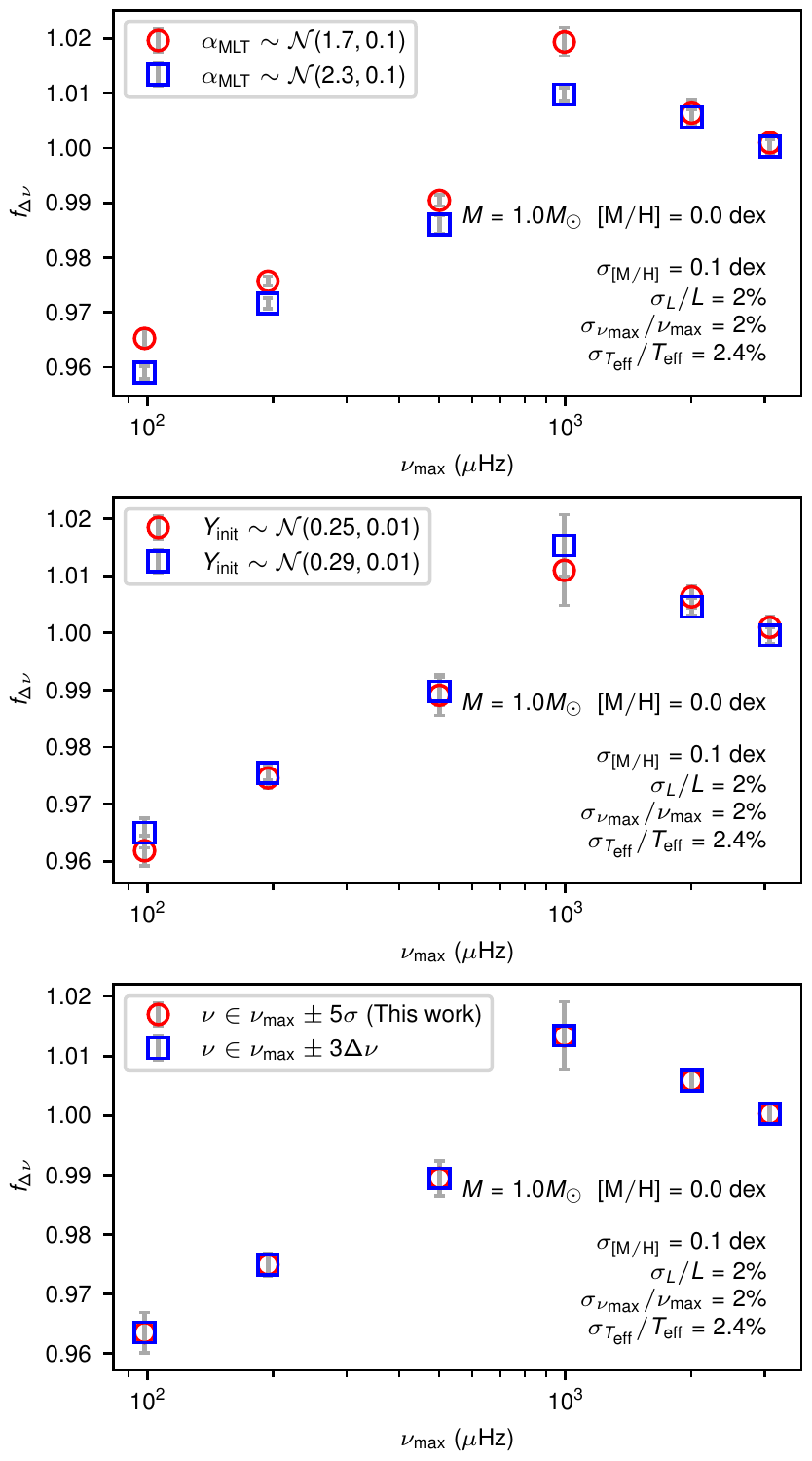}
\caption{Changes of theoretical \fDnu{} due to the choices of priors on \amlt{} and \Yinit{} (top and middle panels) and the ranges of modes to be included when calculating model \Dnu{} (bottom panel). }
\label{fig:fDnu_error}
\end{figure}

We can study two types of uncertainties arising from the calculation of the correction factor \fDnu{}.
The first type of error is due to uncertain stellar physics. 
Using our surface-corrected stellar models, we can quantify the spread of \fDnu{} due to the changes of $Y_{\rm init}$ and $\alpha_{\rm MLT}$, both of which are poorly constrained model parameters.
We generated synthetic stars with the following stellar properties: $L$, \numax{}, \Teff{}, and [M/H], along a 1~\Msolar{}, solar metallicity, pre-RGB-tip evolutionary track from MIST. 
We assumed 2\% observational uncertainties for $L$, 2\% for \numax{}, 2.4\% for \Teff{}, and 0.1 dex for [M/H], according to their typical values \citep[e.g.][]{tayar++2022-error,yuj++2018-16000-rg}.
Then we treated these properties as observational constraints and estimated \fDnu{}, using the fitting routine introduced in Sec.~\ref{subsec:fdnu-results}.
In addition, we assigned Gaussian priors on each model based on its $Y_{\rm init}$ and $\alpha_{\rm MLT}$ values, the results of which are shown in the top and middle panels of Fig.~\ref{fig:fDnu_error}.
Changing \amlt{} clearly has a bigger impact than \Yinit{}. 
The differences on the RGB and the main sequence are smaller than 0.5\%, and in the subgiant phase (\numax{}$\sim1000$~\muHz) they can reach $\sim1$\%.
However, the \fDnu{} itself presents a larger variation than these at most $\sim1$\% uncertainties (see also Fig.~\ref{fig:fDnu}), indicating the necessity of making the correction to the \Dnu{} scaling relation.

The second type of error concerns the range of modes that are used to calculate the theoretical \Dnu{}. We compared two approaches: one using the default range of modes involved in our work, which is $\numax\pm5\sigma$, and another only using the modes in the $\numax\pm3\Dnu$ range. 
From the bottom panel of Fig.~\ref{fig:fDnu_error}, we see negligible differences, indicating that the range of modes considered (as long as larger than 3\Dnu{}) is not a major source of uncertainty.

\bsp	
\label{lastpage}
\end{document}